\documentstyle[prl,aps,multicol]{revtex}
  \renewcommand{\narrowtext}{\begin{multicols}{2} \global\columnwidth20.5pc}
  \renewcommand{\widetext}{\end{multicols} \global\columnwidth42.5pc}
  
  \newcommand{\narrow}{\begin{\flushright}\mbox{\line(0,-1){3}$\! \!$
        \line(1,0){245}} \end{flushright} \narrowtext \noindent}

\input{psfig}

\begin{document}
\bibliographystyle{prsty}
\draft

\title{ Population Inversion Induced by Resonant States in Semiconductors  }

\author{M. A. Odnoblyudov$^{1,2}$,
I. N. Yassievich$^{1,2}$, M. S. Kagan$^3$,
Yu. M. Galperin$^{2,4}$, X. H.  Wang$^1$, and K. A. Chao$^1$}
\address{
$^1$ Department of Theoretical Physics, Lund University,S-223 62,
Lund, Sweden \\ 
$^{2}$A. F. Ioffe  
Physico-Technical  Institute RAS, 194021 St. Petersburg, Russia  \\
$^{3}$Institute for Radioengineering and Electronics RAS, 103907 Moscow, Russia\\
$^{4}$Department of Physics, University of Oslo,  P. O. Box 1048  
Blindern, N 0316 Oslo, Norway
}

\date{\today} \maketitle

\begin{abstract} 
We present a theoretical prediction of a new mechanism for carrier
population inversion in semiconductors under an applied electric field.
The mechanism is originated from a coherent {\it capture-emission} type
inelastic scattering of resonant states. We support our theory with
concrete calculations for shallow acceptor resonant states in strained
$p$-Ge where a lasing in THz frequency region has been recently
observed.
\end{abstract}


\narrowtext
Besides elastic scattering, resonant states also cause a specifically
strong inelastic scattering because electrons can have a finite lifetime
in the resonant state. Consequently, the electron energy is not conserved,
and it can be re-emitted with the energy different from the initial value.
We will prove in this Letter that due to such an effect a novel population
inversion of carrier distribution can be realized in non-equilibrium
semiconductors in the so-called {\em streamer regime}. The streamer regime
emerges in a relatively pure semiconductors at low temperatures where both
impurity scattering and acoustic phonon scattering are weak. Under an
applied electric field a carrier drifts in the momentum space almost scatter
free until its energy reaches the optical phonon energy $\hbar\omega_0$.
Then the carrier emits an optical phonon and return to the low-energy
region~\cite{and}. If there exist a resonant state with energy
$E_0$$<$$\hbar\omega_0$, it acts as a carrier trap with a finite lifetime
which is field-independent. In the energy space one then finds carriers
accumulation in the vicinity of the resonant energy $E_0$.

We will use a general model to analyze this novel mechanism for population
inversion. While our theory applies to a variety of resonant states, in
this Letter we will also demonstrate in quantitative details that our
theory is realistic. For this purpose we will investigate the population
inversion in uniaxially strained $p$-Ge where shallow acceptors induce resonant
states~\cite{str,str1}. It is important to notice that based on such a $p$-Ge
system, a laser operating in terahertz frequency region has been
fabricated recently~\cite{kagan}. By studying the emission lines positions
as functions of applied stress, it was confirmed\cite{str,str1} that the
radiation emission is due to optical transitions between the resonant
states and the localized acceptor (Ga) states. However, the mechanism of
lasing has remained a puzzle. Our theory will explain the physical origin
of the population inversion which leads to the lasing. 

We consider a model system of charge carriers interacting with optical
phonons of energy $\hbar\omega_0$ under a strong external electric field
$\cal E$ along the $z$ axis. In this case an electron acquire energy from
the electric field and {\it drift} in the $\bf k$-space until its energy
exceeds $\hbar\omega_0$. Then the electron emits an optical phonon and
returns to the region $k$$\approx$0. Thus, our model includes a {\it drain}
$D$ at $E_{\bf k}$=$\hbar\omega_0$ and a {\it source} $S$ at small $k$. 
In the absence of resonant scattering by impurities, for carrier kinetic
energy $E_{\bf k}$$\le$$\hbar\omega_0$, the carrier distribution function
$f_{\bf k}$ can be obtained from the kinetic equation~\cite{and})
\begin{equation} \label{e_1}
{\partial f_{\bf k}\over\partial t}+
{{\rm e\cal{E}}\over \hbar}{\partial f_{\bf k}\over\partial{\rm k_z}}=
S- D \, . 
\end{equation}

For our problem we can well approximate the drain $D$ by a black-wall
boundary condition that $f_{\bf k}$=0 for $E_{\bf k}$$\ge$$\hbar\omega_0$.
The source intensity $S$ is determined by the ${\bf k}$-space particle flow
with energy $E_{\bf k}$=$\hbar\omega_0$.
In the following we use the expression for $S$ as~\cite{and}
\begin{equation} \label{source}
S=S_0 (t) \, \Theta(\epsilon_0 - E_{\bf k})\, .
\end{equation} 
Here 
\begin{equation} \label{epsilon_0}
\epsilon_0 = \left( \frac{2}{9}\right)^{1/3}
\left(\frac{\omega_0}{\nu_A} \right)^{2/3} \frac{(e{\cal E}
\hbar)^{2/3}}{m_z^{1/3}}\, , 
\end{equation}
with $m_z$ being the effective mass along $z$-axis. The frequency $\nu_A$
is related to the rate $\nu_A\sqrt{(E_{\bf k}/\hbar\omega_0)-1}$ of optical
phonon emission by the carriers with $E_{\bf k}$$>$$\hbar\omega_0$. The
source amplitude $S_0(t)$ is determined from the condition of particle 
flow conservation in the $\bf k$-space, namely the flow out of the
surface $E_{\bf k}$=$\hbar\omega_0$ at time $t$ returns evenly back into
the region $E_{\bf k}$$<$$\epsilon_0$,
\begin{equation} \label{source1}
S_0 (t) = \frac{e}{\hbar}\frac{ \int (\bbox{\cal E}\cdot d{\bf S}) \,
f_{\bf k}(t)}{\int d^3 k \, \Theta (\epsilon_0 - E_{\bf k})} \, .
\end{equation}
The integration is performed over the surface defined by the equation
$E_{\bf k}$=$\hbar \omega_0$.

The stationary solution of Eq.~(\ref{e_1})
(the so-called {\em streamer}) is that $f_{\bf k}$ is almost constant
if ${\bf k}$ lies in a cylinder, and $f_{\bf k}$=0 otherwise. 
This cylinder in $\bf k$-space is determined by
\[
0< k_z \le \sqrt{2m_z\omega_0/\hbar^2} \, , \quad 
k_\perp \le \sqrt{2m_\perp \epsilon_0/\hbar^2} ,
\]
where $m_\perp$ is the transverse component of the effective mass tensor. 

When impurities induce resonant states with complex energy
$E_0$+$i\Gamma/2$, charge carriers may be trapped at energy
$E_0$$<$$\hbar\omega_0$ for a time interval $\sim\hbar/\Gamma$. As a result,
a maximum of the non-equilibrium distribution function is formed around
the energy $E_0$. This is just the population inversion. To take into
account the resonant scattering by impurities of concentration $N_i$, at
the right hand side of Eq.~(\ref{e_1}), we should add the impurity
collision integral $I$,
\begin{eqnarray} \label{ci}
I &=& N_i V \sum_{\bf k^\prime}\left[f_{{\bf k}^\prime}W_{{\bf
kk}^\prime} - f_{\bf k}W_{\bf k^\prime k}\right] 
\nonumber \\
&& \qquad + N_i V \left(W_{r{\bf k}}f_r-W_{{\bf k}r}f_{\bf k}\right) \, ,
\end{eqnarray}
where $V$ is a normalization volume, and $f_r$ is the resonant state
population which satisfies the kinetic equation
\begin{equation} \label{e_3}
{\partial f_r\over\partial t}=\sum_{\bf k^\prime}
\left[W_{{\bf k^\prime}r}f_{\bf k^\prime}-W_{r{\bf k^\prime}}f_r\right].
\label{kin_sys}
\end{equation}
The first term in Eq.~(\ref{ci}) represents the elastic scattering, while
the second one describes the coherent {\it capture} and {\it re-emission}
by the resonate state. The sum $\sum_{{\bf k}'} W_{r{\bf k}'}$ is the
total escape rate from the resonant state, and is equal to $\Gamma/\hbar$.

The transition probabilities $W_{{\bf kk}'}$ and $W_{{\bf k}r}$ can be
expressed through the respective scattering amplitude $t_{{\bf kk}'}$ and
transition amplitude $t_{{\bf k}r}$ as
\begin{eqnarray}
W_{{\bf kk}'} &=&
\frac{2\pi}{\hbar}|t_{{\bf kk}'}|^2\,\delta(E_{\bf k} - E_{{\bf k}'})
\, , \\
W_{{\bf k}r} &=&
\frac{2 }{\hbar}|t_{{\bf k}r}|^2 \,
\frac{\Gamma/2}{(E_{\bf k} -E_0)^2 + \Gamma^2/4}\, .
\end{eqnarray}
$t_{\bf{kk}'}$ and $t_{{\bf k}r}$ can be calculated using the Dirac
approach~\cite{dirac} to the scattering problem at a resonant state.
According to this approach, via hybridization with extended states, the
wave function $\varphi({\bf r})$ of the {\it bare} localized impurity state
develops into the resonant state. The general form of the scattering
state is
\begin{eqnarray}
\Psi_{\bf k} ({\bf r}) &=&
\frac{1}{\sqrt{V}}e^{i {\bf kr}} +
\frac{t_{\bf{k}r}}{E_{\bf k} - E_0 +
i \Gamma/2} \,\varphi ({\bf r}) \nonumber \\ && +
\sum_{{\bf k}'} \frac{t_{\bf{kk}'}}{E_{\bf k} - E_{{\bf k}'} +
i \gamma}\, \frac{1}{\sqrt{V}}e^{i {\bf k'r}} \, ,
\quad \gamma \rightarrow +0 \, .
\end{eqnarray}

Finally, we will add the normalization condition
\begin{equation} \label{e_2}
\sum_{\bf k} f_{\bf k} + N_iV f_r = nV
\end{equation}
where $n$ is the total electron concentration, and solve all these
equations selfconsistently.

The above theory is general for different types of resonant states. To
demonstrate explicitly the population inversion predicted by our theory,
we will investigate quantitatively the resonant states induced by shallow
acceptors in uniaxially strained $p$-Ge. 

In cubic semiconductors with symmetry group $O_h$ the shallow acceptor
wave functions $\phi^{(M)}({\bf r})$ are 4-fold degenerate with the total
angular momentum projections $M$=$\pm 1/2$ and $\pm 3/2$. The ground state has
$\Gamma^+_8$ symmetry. Under a uniaxial strain the valence band top splits
into two doubly degenerate energy levels. The corresponding two sets of
wave functions transform according to $\Gamma^+_6$ and $\Gamma^+_7$
representations if the stress is along the [001] direction, and according
to $\Gamma^+_4$ and $\Gamma^+_{5+6}$ representations if the stress is along
the [111] direction. The ground state acceptor wave functions split in the
same way and can be classified by the total angular moment projections
($M$=$\pm 1/2$ or $M$=$\pm 3/2$) along the stress direction. In our
calculations, we will use the spherical approximation for the Luttinger
Hamiltonian (LH)~\cite{bl}. In the limit of large strain such that the
splitting $E_d$ at the top of subband exceeds the Coulomb energy, the LH
can be treated with a quasi-diagonal approximation, and is represented by
two 2$\times$2 blocks. The states in each subband can now be classified by
the projections ($m = \pm 1/2$ or $\pm 3/2$) of the {\em hole spin}  on the
stress axis~\cite{bp}.    

\begin{figure}[h]
\centerline{
\psfig{figure=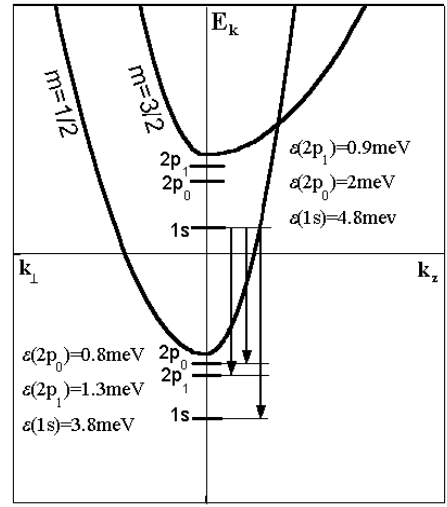,width=6cm}
}
\caption{The acceptor levels diagram for uniaxially strained Ge in the
large strain limit along [001]. The optical transitions observed in the
lasing{\protect\cite{kagan}} are indicated by arrows.
\label{fig1}}
\end{figure}

In this way, in each of doubly-degenerate subbands, we obtain both extended
states and the localized Coulomb states below the subband. The
corresponding energy levels and wave functions are calculated following the
variational procedure in Ref.~\onlinecite{str,str1}. For a large
strain, the energy levels are shown in Fig.~\ref{fig1}.

The {\it off-diagonal} terms of LH mix the states belonging to different
subbands. Treating these terms as perturbation in the Dirac
approach~\cite{dirac} the hybridization of local and extended states
yields the resonant state with energy $E_0$ and level width $\Gamma$.
Along the Dirac approach one also obtains the scattering amplitudes
$t_{{\bf kk}'}^{mm'}$ ($m,m'$=$\pm 1/2$) and the transition amplitudes
$t_{{\bf k}r}^{mm'}$ ($m$=$\pm 3/2$ and $m'$=$\pm 1/2$). The explicit
expressions of these amplitudes, as well as the details of deriving them
will be published elsewhere. 

We are ready to use these scattering and transition amplitudes to solve
the set of kinetic equations (\ref{e_1})-(\ref{e_3}). However, for $p$-Ge
the kinetic equations can be simplified. Let $\tau_{\cal E}$ be the
transient time during which the stationary non-equilibrium distribution
is established. $\tau_{\cal E}$ depends on both the electric field $\cal E$
and the impurity concentration $N_i$. As will be shown below, for
$p$-Ge, there exists rather large region of electric fields and
impurity concentrations where
$\tau_{\cal E}$ is much larger than the lifetime of the resonant state
$\hbar/\Gamma$. In this case we can set the left hand side of
Eq.~(\ref{e_3}) to zero, and so the occupation of the quasi-local states,
$f_r$, follows adiabatically the distribution function $f_{\bf k}$ of the
extended states. Since the localized and the extended states are doubly
degenerate, we have $f_r^{+3/2}$=$f_r^{-3/2}$$\equiv$$f_r$ and
$f_{\bf k}^{+1/2}$=$f_{\bf k}^{-1/2}$$\equiv$$f_{\bf k}$. If we define
$|t_{{\bf k}r}|^2 \equiv \left|t_{{\bf k}r}^{1/2,3/2}\right|^2+\left|t_{{\bf
k}r}^{1/2,-3/2}\right|^2$, we obtain 
\begin{equation} \label{e_4}
f_r = \sum_{{\bf k}} \frac{|t_{{\bf k}r}|^2f_{\bf k}}{(E_{{\bf k}} -
E_0)^2 + \Gamma^2/4} \, .
\end{equation}   
Substituting Eq.~(\ref{e_4}) into Eqs. (\ref{e_1}) and (\ref{ci}), we arrive
at the kinetic equation for $ f_{\bf k}$
\begin{eqnarray} \label{e_5}
&& {\partial f_{\bf k}\over\partial t} +
{{\rm e\cal{E}}\over \hbar}{\partial f_{\bf k}\over\partial{\rm k_z}} =
\frac{2 \pi N_iV}{\hbar} \sum_{{\bf k}'}|t_{{\bf kk}'}|^2 
\delta (E_{\bf k} -E_{{\bf k}'}) (f_{{\bf k}'} - f_{{\bf k}})
\nonumber \\
&& \ + \frac{N_i V |t_{{\bf k}r}|^2 \Gamma}{\hbar[(E_{\bf k} -E_0)^2 +
\Gamma^2/4]}\left[\sum_{{\bf k}'}
\frac{|t_{{\bf k}'r}|^2\, f_{{\bf k}'}}{(E_{\bf k} -E_0)^2 +
\Gamma^2/4} - f_{\bf k}\right] \nonumber \\
&& \quad  + S_0 (t) \, \Theta(\epsilon_0- E_{\bf k}) \, ,
\end{eqnarray}
where $|t_{{\bf kk}'}|^2$$\equiv$$\left|t_{{\bf kk}'}^{1/2,1/2}\right|^2+
\left|t_{{\bf kk}'}^{1/2,-1/2}\right|^2$. The boundary condition for the
above kinetic equation is $f_{\bf k}$=0 at $E_{\bf k}$=$\hbar\omega_0$. 
The source $S_0 (t)$ and the energy $\epsilon_0$ are given by Eqs.
(\ref{source1}) and (\ref{epsilon_0}), respectively.

We will solve Eq.~(\ref{e_5}) numerically as a non-stationary equation. We
start from some initial distribution and then follow the evolution of the
distribution function until it reaches a stationary one. In this way, the
final stationary distribution and the transient time $\tau_{\cal E}$ are
obtained for various values of applied electric field and impurity
concentration. In our numerical calculation, the values of material
parameters for $p$-Ge are Luttinger parameters $\gamma_1$=13.38,
$\gamma_2$=4.24, $\gamma_3$=5.69, $\hbar \omega_0 = 36$~meV, and the
characteristic frequency 
$\nu_A$=5$\times$10$^{12}$ s$^{-1}$. 
\begin{figure}[h]
\centerline{
\psfig{figure=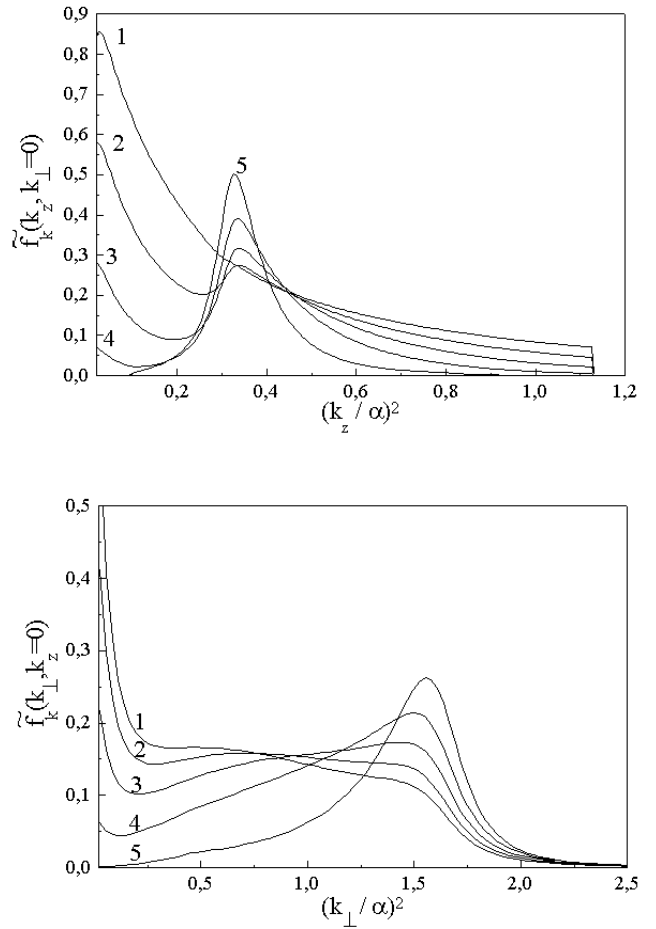,width=8.5cm}
}
\caption{
Normalized distribution function
$(\alpha/2\pi)^3 n^{-1}f_{\bf k}$ as a function of squared normalized momentum
$(k_z/\alpha)^2$ (upper panel) and $(k_{\perp}/\alpha)^2$ (lower panel) in a
uniaxially strained $p$-Ge under an electric field ${\cal E}=100$~V/cm
along the $z$ axis. Various curves are for impurity concentration
$N_i$=1 (curve 1), 1.1 
(curve 2), 1.2 (curve 3), 1.4 (curve 4), and 1.5 (curve 1), in units
10$^{15}$~cm$^{-3}$.
\label{fig2}} 
\end{figure}
We will set the pressure at 5 kbar and the electric field at
${\cal E}$=100 V/cm along [111]. In this case, the resonant level is at
the energy $E_0=10$~meV and has a width $\Gamma=2$~meV. Here we have
adopted the approximation of a single hole band, and we are aware of the
fact that at a pressure of 5 kbar the inter-hole-band splitting
$E_d$=20 meV is less than $\hbar\omega_0$. One can show that including the
second band will only change the tail of the distribution, and such change
is not important for the problem under consideration. From
Eq~(\ref{epsilon_0}) we obtain the source width $\epsilon_0$=4.25 meV.
If we define $\alpha$=$\sqrt{2m_0E_0/\hbar^2 \gamma_1}$ as the unit for
wave vector, 
where $m_0$ is the free electron mass, our calculated normalized
distribution functions are shown in Fig.~\ref{fig2} as functions of
normalized $k_z/\alpha$ (upper panel) and of normalized $k_\perp/\alpha$
(lower panel). The curves in Fig.~2 are for impurity concentration $N_i$=1
(curve 1), 1.1 (curve 2), 1.2 (curve 3), 1.4 (curve 4), and 1.5 (curve 1),
in units 10$^{15}$ cm$^{-3}$. While the distribution function peaks are
centered at the resonant energy $E_0$, their corresponding peak positions
in the upper panel are different from those in the lower panel. This is
due the anisotropy of the effective mass: $m_z$=0.04$m_0$ and
$m_\perp$=0.13$m_0$. For the results shown in Fig.~\ref{fig2}, the
transient time is about $\tau_{\cal E}$$\approx$10$^{-11}$~s, which is
much longer than the lifetime $\hbar/\Gamma$$\approx$2$\times$10$^{-13}$~s.
Consequently, our calculation based on the condition that
$\Gamma\tau_{\cal E}$$\gg$$\hbar$ is selfconsistent.
\begin{figure}
\centerline{
\psfig{figure=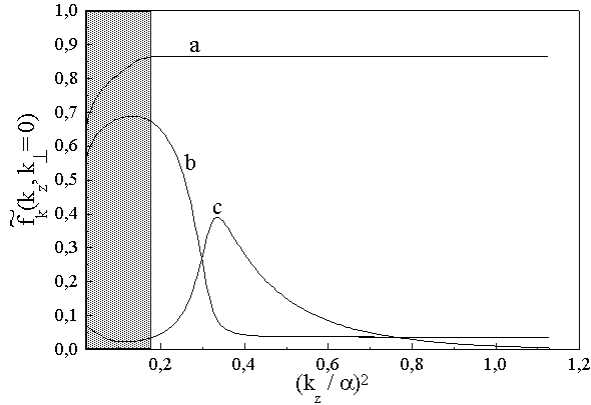,width=8.3cm}
}
\caption{Normalized distribution function as a function of squared normalized
momentum $(k_z/\alpha)^2$ for impurity concentration $N_i=5 \times
10^{15}$~cm$^{-3}$. The curves are: (a) no resonant scattering, (b)
include only 
elastic scattering from resonant states, and (c) include full resonant
scattering.\label{fig3}}
\end{figure}
In order to demonstrate precisely which scattering process is responsible
to the population inversion, we have used the same values of material
parameters and same electrical field strength to calculate the distribution
functions for 3 cases, and the results are shown in Fig.~\ref{fig3}. In the
absence of resonant states, the distribution function follows curve (a).
When elastic scattering due to resonant states is taken into account, the
result change into the step-like curve (b). Finally, by adding the
inelastic scattering process, a peak emerges in curve (c). It is then clear
that the population inversion is induced by the {\it capture-emission}
component. The population accumulated in resonant states is controlled
by the distribution function in the continuous spectrum, as indicated by
Eq.~(\ref{e_4}). At the same time, the non-equilibrium population in the
localized states can be only less than that in the low-energy continuous
states. Consequently, under the conditions imposed on our calculation, the
intra-center population is also inverted. It has been demonstrated in
Ref.~\onlinecite{str,str1} that the lasing in uniaxially strained Ge:Ga is 
connected to the transitions shown by arrows in Fig.~\ref{fig1}. Among
these transitions, besides the allowed intra-center optical transitions
between the lowest resonant $s$-type state and excited localized $p$-type
states, there  is also transition corresponding to the {\it forbidden}
$s$-type to $s$-type. The observed {\it forbidden} transition is due to
the accumulation of the continuous spectrum carriers at the energy around
$E_0$. These states are almost plane waves and so turn the {\it forbidden}
process into an allowed one. The sharp emission line in the observed
optical spectrum proves the existence of the peak in the distribution
function $f_{\bf k}$.  Thus, the resonant-states-induced population
inversion predicted by our theory explains the origin of lasing in
terahertz frequency range observed in these strained Ge:Ga samples.

In conclusion, we have predicted that resonant states can produce a
population inversion in the carrier distribution function in strained
semiconductors under an external electric field. Our theoretical
prediction is confirmed by concrete calculations for strained $p$-Ge,
where resonant states give rise to the lasing observed in THz frequency
region. We believe that the proposed mechanism for population inversion is
rather general since resonant states can be created by various means.

We thank V. I. Perel and A. A. Andronov for useful discussions. Financial
supports from Swedish Natural Science Research Council Grant No. \"O-AH/KG
03996-322 and 03996-321, by Nordic Academy for Advanced Study (NorFA)
Grant No. 98.55.002-O, and by Russian Foundation for Basic Research
Grant No. 98-02-18268 and 97-02-16820 are acknowledged.


\widetext
\end{document}